\documentclass[twocolumn,twoside,slac_two]{revtex4}
\usepackage{graphicx}
\usepackage{fancyhdr}
\def\xspace     {\,}
\def\to         {\ensuremath{\rightarrow}\xspace}
\def\etal       {{\it et~al.}}
\def\BR         {{\ensuremath{\cal B}\xspace}}

\def\liferatio  {\ensuremath{\frac{\tau_{\Bp}}{\tau_{\Bz}}}\xspace}
\def\PM #1 #2   {^{+#1}_{-#2}{}}
\def\pvecB      {\vec{p}{}^{\;*}_B}
\def\Ebeam      {\ensuremath{E^*_{beam}}}
\def\Afb        {\ensuremath{A_{FB}}}

\usepackage{relsize}
\def\babar{\mbox{\slshape B\kern-0.1em{\smaller A}\kern-0.1em
    B\kern-0.1em{\smaller A\kern-0.2em R}}}

\def\Kstar   {\ensuremath{K^*}\xspace}
\def\Bz      {\ensuremath{B^0}\xspace}
\def\Bp      {\ensuremath{B^+}\xspace}
\mathchardef\Lambda="7103
\def\Lambar  {\kern 0.2em\overline{\kern -0.2em\Lambda\kern 0.05em}\kern-0.05em{}\xspace}
\def\pip     {\ensuremath{\pi^+}\xspace}
\def\piz     {\ensuremath{\pi^0}\xspace}
\def\etapr   {\ensuremath{\eta^{\prime}}\xspace}

\def\bsg        {\ensuremath{b \to s \gamma}\xspace}
\def\bxsg       {\ensuremath{B \to X_{s} \gamma}\xspace}
\def\bksg       {\ensuremath{B \to \Kstar \gamma}\xspace}
\def\bks892g    {\ensuremath{B \to K^*(892) \gamma}\xspace}
\def\bdg        {\ensuremath{b \to d \gamma}\xspace}
\def\bxdg       {\ensuremath{B \to X_{d} \gamma}\xspace}
\def\brpg       {\ensuremath{\Bp \to \rho^+ \gamma}\xspace}
\def\brzg       {\ensuremath{\Bz \to \rho^0 \gamma}\xspace}
\def\bog        {\ensuremath{\Bz \to \omega \gamma}\xspace}
\def\brog       {\ensuremath{B \to (\rho/\omega) \gamma}\xspace}
\def\bsll       {\ensuremath{b \to s \ell \ell}\xspace}
\def\bxsll      {\ensuremath{B \to X_{s} \ell \ell}\xspace}
\def\bkll       {\ensuremath{B \to K \ell \ell}\xspace}
\def\bksll      {\ensuremath{B \to \Kstar \ell \ell}\xspace}
\def\bxclnu     {\ensuremath{B \to X_{c} \ell \nu}\xspace}

\def\invfb {\ensuremath{\mbox{\,fb}^{-1}}\xspace}
\def\invab {\ensuremath{\mbox{\,ab}^{-1}}\xspace}
\newcommand{\gev}{\ensuremath{\mathrm{\,Ge\kern -0.1em V}}\xspace}
\newcommand{\gevns}{\ensuremath{\mathrm{\,Ge\kern -0.1em V}}}

\def\mum   {\ensuremath{{\,\mu\rm m}}\xspace}

\def\app   #1 #2 #3 {Acta.~Phys.~Polon.~{\bf#1},\ #2 (#3)}
\def\ijm   #1 #2 #3 {Int.~J.~Mod.~Phys.~A~{\bf#1},\ #2 (#3)}
\def\nima  #1 #2 #3 {Nucl.~Instrum.~Methods~A~{\bf#1},\ #2 (#3)}
\def\npb   #1 #2 #3 {Nucl.~Phys.~B~{\bf#1},\ #2 (#3)}
\def\plb   #1 #2 #3 {Phys.~Lett.~B~{\bf#1},\ #2 (#3)}
\def\prd   #1 #2 #3 {Phys.~Rev.~D~{\bf#1},\ #2 (#3)}
\def\prl   #1 #2 #3 {Phys.~Rev.~Lett.~{\bf#1},\ #2 (#3)}
\def\rmp   #1 #2 #3 {Rev.~Mod.~Phys.~{\bf#1},\ #2 (#3)}

\begin{document}

\title{Radiative Penguin Decays at the B Factories}

\author{V. E. \"Ozcan}
\affiliation{University College London, WC1E 6BT, United Kingdom}
\begin{abstract}
An overview of the measurements of 
$b\rightarrow s\gamma$, $b\rightarrow d\gamma$ and $b\rightarrow s\ell\ell$ 
penguin transitions at the $B$~Factories is presented.
\end{abstract}

\maketitle

\thispagestyle{fancy}


\section{Introduction}
The flavor-changing neutral current transitions $b \to s$ and $b \to d$
cannot proceed at the tree level in the Standard Model~(SM). Radiative
decays such as \bxsg, \bxdg\ and \bxsll\ hence proceed primarily via one-loop
diagrams known as penguin and box diagrams. Because of this, these processes
are rare in the SM, with branching ratios on the order of $10^{-4}$ to
$10^{-6}$. However, for the same reason, these channels are very sensitive
to new physics beyond the SM.

In the past few years, the $B$ Factory detectors, Belle at the KEKB collider at
KEK, Japan and \babar\ at the PEP-II collider at SLAC, U.S.A., have recorded
a total of more than 0.9\invab of data, making it possible to study these decays
extensively.  Both experiments benefit from high-granularity electromagnetic calorimetry
with excellent angular and energy resolution ($\sim\!3\%$ for 1\gev photons), silicon
vertex detectors providing position resolution on the order of 100\mum along the beam
axis and $K$/$\pi$ separation for momenta up to 3-4\gevns. Details of the
detectors can be found in references~\cite{bib:detectors}.

The relevant theoretical framework for all the decays reviewed in this paper
is the operator product expansion~(OPE)~\cite{bib:hurth_summary}, with which the
heavy particles (the gauge bosons and the top quark in the SM) are integrated out to
obtain an effective low-energy theory with five quarks.  The long-distance
contributions are captured in operator matrix elements, whereas
the short-distance physics is described by the effective {\emph coupling constants},
known as Wilson coefficients.  Physics beyond the SM can manifest
in new operators, or in vastly different values for the Wilson coefficients of
the existing operators, compared to the SM predictions.

The OPE framework provides parton-level predictions, so the connection to the
experimental observables require further non-perturbative corrections.  Form factors,
calculations of which introduce significant theoretical uncertainties, are
needed to study exclusive final states~(\bksg, \bkll,~etc.).  On the other hand,
for inclusive decays, heavy-quark expansion~(HQE) and quark-hadron duality keep
non-perturbative corrections under control.
This allows more direct
interpretations of the measurements of the inclusive decay rates.  However,
exclusive decays, being experimentally cleaner, are observed earlier
and measured with better precision.  Ratios of exclusive-decay observables,
in which the form-factor corrections partially or completely cancel, can
also provide complimentary information to that obtained from inclusive decays.

\section{\boldmath\bsg}
\subsection{Exclusive \boldmath\bsg}

\begin{figure}[h]
\centering
\includegraphics[bb = 0 0 455 740, width=80mm]{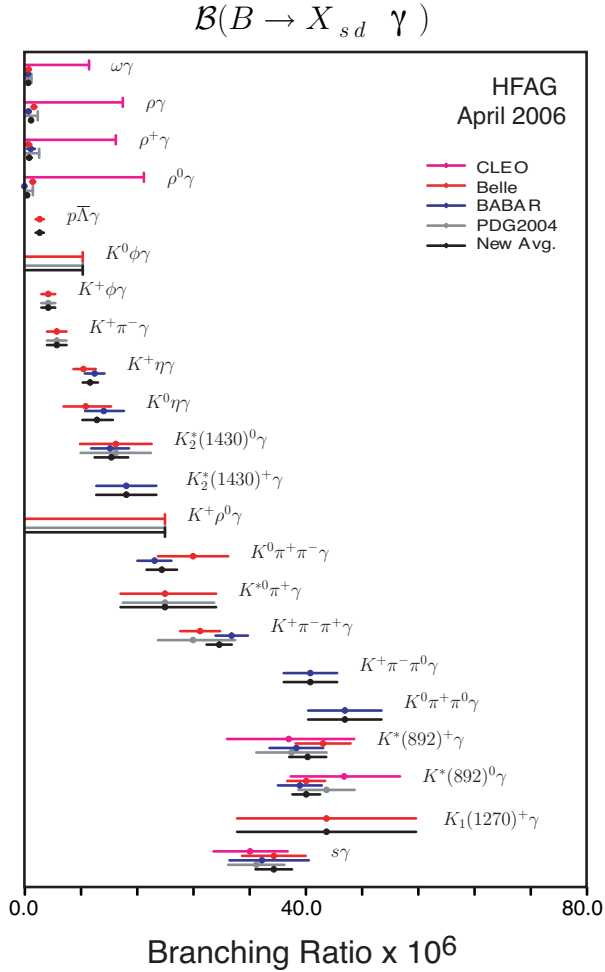}
\caption{Branching fractions for $b\to s/d$ photon
penguins~\cite{bib:HFAG06}.  The value for inclusive $s\gamma$ has been
reduced by an order of magnitude to fit it in the figure.}
\label{fig:radpengs}
\end{figure}

Since its first observation in the exclusive mode \bks892g \ in
1993~\cite{bib:cleofirst}, \bsg\ has been the most extensively studied
radiative penguin process.  With the large data samples of the $B$ Factories,
many other exclusive modes have since become accessible and so far the
branching fractions of nine charged and six neutral final states have been
measured at $3\sigma$ or better significance
(Fig.~\ref{fig:radpengs})~\cite{bib:HFAG06}.  For some of these modes,
measurements of the charge asymmetry and the time-dependent $CP$ violation
have also been done~\cite{bib:kakuno_fpcp}; the results are so far consistent with
the predictions from the SM.

It is worth highligthing a few of the most recent results.  The evidence
for the first baryonic final state, $\Bp\to p\Lambar\gamma$, has been reported
by the Belle collaboration in 2004~\cite{bib:belle_plamg}.  While this decay has currently
the smallest branching fraction ($(2.16^{+0.58}_{-0.53}\pm 0.20)\times 10^{-6}$)
among all the \bsg\ final states measured at $3\sigma$ or better significance, its
penguin nature becomes apparent due to the stringent upper limit on the two-body
decay $\Bp\to p\Lambar$~\cite{bib:belle_plam}.

More recently, the observation by the Belle Collaboration
of $\Bp\to K_1^+(1270) \gamma$, the first
radiative $B$-meson decay involving an axial-vector resonance,
with an unexpectedly high branching fraction ($(43\pm9 \pm9)\times 10^{-6}$) 
has attracted some interest~\cite{bib:belle_k1270g}.  Inclusive measurements of
four different $K\pi\pi\gamma$ final states by the \babar\ Collaboration yield
similarly high branching fractions~\cite{bib:babar_kppg}.  With more
statistics, the decays $B\to K\pip\piz\gamma$ will permit the measurement of
the photon polarization, which is almost completely left-handed in the
SM~\cite{bib:kppg_polarization}.

Finally, the most recent development in the exclusive \bsg\ front is the
confirmation by \babar\ of Belle's earlier measurement of the
$B\to K\eta\gamma$ decay~\cite{bib:ketag}.  Both charged and neutral modes are now
established at better than $5\sigma$ significance.  Additionally
\babar\ has also obtained the first upper limits on the decay
$B\to K\etapr\gamma$.  These upper limits (at~90\%~C.L.) and the new combined
averages for the $K\eta\gamma$ final states by the Heavy Flavor
Averaging Group~(HFAG) are given in Eq.~\ref{eq:ketag}:
\begin{eqnarray}
\BR(\Bp\to K^{+}\eta\gamma) & = & (9.3 \pm 1.1)\times 10^{-6}\,,  \nonumber \\
\BR(\Bz\to K^{0}\eta\gamma) & = & (10.3^{+2.3}_{-2.1})\times 10^{-6}\,,  \nonumber \\
\BR(\Bp\to K^{+}\etapr\gamma) & < & 4.2\times 10^{-6} \,,\nonumber \\
\BR(\Bz\to K^{0}\etapr\gamma) & < & 6.6\times 10^{-6} \,.
\label{eq:ketag}
\end{eqnarray}
These measurements will be useful in reducing the systematic uncertainties
in the inclusive measurements of \bxsg, and in searches for higher-mass \Kstar
resonances.  Moreover, it is possible to measure time-dependent $CP$ asymmetry
in the neutral final states.

\subsection{Inclusive \boldmath\bsg}

Inclusive measurement of the \bxsg total rate,\footnote{Both within the theory
community and among the experimentalists, the \bxsg total rate usually refers to
the decay rate for photon energy above 1.6\gevns~\cite{bib:BurasMisiak}.}
where $X_s$ refers to any hadronic system with unit strangeness,
gives one of the most widely quoted constraints on physics beyond the
SM.  Additionally, the shape of the photon energy spectrum, which is
mostly insensitive to new physics, provides essential input for measuring
Cabibbo-Kobayashi-Maskawa (CKM) matrix elements $V_{cb}$ and $V_{ub}$
from semileptonic $B$-meson decays.  Therefore,
this decay has been measured multiple times using independent techniques.

The current experimental world average reported by HFAG, 
$(3.55\pm 0.24^{+0.09}_{-0.10}\pm 0.03)\times 10^{-4}$, where the errors are
combined statistical and systematic, systematic due to the shape function, and
systematic due to the subtraction of $d\gamma$ contribution, incorporates five
branching fraction
measurements from CLEO, Belle and \babar\ Collaborations~\cite{bib:HFAG06}.  All
five measurements are in very good agreement with each other and with the
SM prediction of $(3.61^{+0.37}_{-0.49})\times 10^{-4}$, computed at the
next-to-leading-log precision~\cite{bib:Hurthfinal}.  This remarkable agreement
implies strong constraints on various models of new physics.  For example,
in type-II two-Higgs-doublet model, which represents a good approximation for
gauge-mediated supersymmetric models with large $\tan\beta$, where the charged
Higgs contribution dominates the chargino contribution, the current experimental
value implies a minimimum charged-Higgs-boson mass of $350\gev$~\cite{bib:babarphysreach}.

The moments obtained from the photon energy spectra are consistent
among the different measurements and with the HQE predictions
based on inputs from the hadron-mass and lepton-energy spectra measured
in \bxclnu\ decays.  A fit to all the available data from
\bxsg\ and \bxclnu\ decays has yielded precise values for the
HQE parameters $m_b$ and $\mu^2_\pi$, as shown in Table~\ref{tab:mbmupi2}.
\begin{table}[h]
\begin{center}
\caption{Heavy-quark expansion parameters $m_b$ and $\mu^2_\pi$, and their
correlation $\rho$, obtained from the combined fit to \bxsg and \bxclnu
data in two HQE schemes~\cite{bib:HQEfit}.}
\begin{tabular}{|l|c|c|c|}
\hline Scheme  & $m_b (\gevns)$  & $\mu^2_\pi (\gevns^2)$ & $\rho$  \\
\hline Kinetic & $4.590\pm0.039$ & $0.401\pm0.040$ & $-0.39$ \\
\hline Shape Function & $4.604\pm0.038$ & $0.189\pm0.038$ & $-0.23$ \\
\hline
\end{tabular}
\label{tab:mbmupi2}
\end{center}
\end{table}
Further details on the derivation of these parameters and the extraction
of $|V_{cb}|$ and $|V_{ub}|$ can be found elsewhere in the FPCP conference
proceedings~\cite{bib:fpcp_inclsemilept}.

\section{\boldmath\bdg}

Compared to \bsg, the amplitude for the process \bdg\ is suppressed
by the ratio of the CKM matrix elements $|V_{td}|/|V_{ts}|$.  Therefore
it is experimentally more difficult to observe than \bsg: the branching
fractions for even
the most prominant exclusive final states are expected to be on the order
of $10^{-6}$.  Furthermore, misidentified \bsg\ decays add up to the 
other substantial backgrounds (mostly high-energy photons produced through
$\piz/\eta\to\gamma\gamma$ decays or via initial-state radiation in
$e^+e^-\to q\bar{q}$ continuum events, where $q=u,d,s,c$) that are 
common to the measurements of all the radiative penguin decays.

\begin{figure}[h]
\centering
\includegraphics[width=80mm]{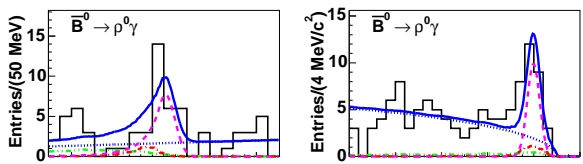}
\caption{Projections of the fit results to $M_{bc}$ and $\Delta E$
distributions for the Belle measurement of the decay mode \brzg.
Curves show the signal (dashed), continuum (dotted), \bks892g (dot-dashed),
other-$B$-decay background (dot-dot-dashed) components, and the total fit
result (solid)~\cite{bib:belle_bdg}.}
\label{fig:belle_bdg}
\end{figure}

The search for \bdg\ has been carried out by CLEO, \babar\ and Belle
collaborations in one charged (\brpg) and three neutral
($\Bz\to\rho^0/\omega/\phi\,\,\gamma$) exclusive decays.  It is customary
to combine the charged and neutral $\rho$ and $\omega$ modes using
isospin relations to obtain an average 
branching fraction defined by:
$\BR(\brog) = \frac{1}{2} \{ \BR(\brpg) + \liferatio
\left[ \BR(\brzg) + \BR(\bog) \right] \}$,
where \liferatio is the ratio of the $B$-meson lifetimes.  The branching
fractions for each mode from Belle and \babar\ are
listed in Table~\ref{tab:bdg}.  For the $\rho$ and $\omega$ modes,
center values are given for comparison, even in the absence of any significant signal.
\begin{table}[h]
\begin{center}
\caption{Branching fraction and significance for each \bdg\ mode
from Belle~\cite{bib:belle_bdg} and \babar~\cite{bib:babar_bdg}.} 
\begin{tabular}{|l|c|c|c|c|} \hline
Mode  & $\BR(10^{-6})$ & Sig. & $\BR(10^{-6})$ & Sig.  \\ \hline
\brpg & $0.55\PM 0.42 0.36 \PM 0.09 0.08 $ & 1.6 & $0.9\,\PM 0.6 0.5 \pm 0.1$ & 1.9 \\
\brzg & $1.25\PM 0.37 0.33 \PM 0.07 0.06 $ & 5.2 & $0.0\pm 0.2 \pm 0.1$ & 0.0 \\
\bog  & $0.56\PM 0.34 0.27 \PM 0.05 0.10 $ & 2.3 & $0.5\pm 0.3 \pm 0.1$ & 1.5 \\
\brog & $1.32\PM 0.34 0.31 \PM 0.10 0.09 $ & 5.1 & $0.6\pm 0.3 \pm 0.1$ & 2.1 \\ \hline
$B\to\phi\gamma$ & \multicolumn{2}{|c|}{---} & \multicolumn{2}{|c|}{$<0.85$ (90\% C.L.)} \\
\hline
\end{tabular}
\label{tab:bdg}
\end{center}
\end{table}
While Belle and \babar\ results do not agree well with each other for the decay \brzg,
the clear signal in Belle measurement of this mode gives the
first observation of the \bdg
process.  Figure~\ref{fig:belle_bdg} shows the beam-energy constrained
mass $M_{bc}= \sqrt{ (\Ebeam/c^2)^2 - |\pvecB/c|^{2}}$, and the energy
difference $\Delta E = E^*_B - \Ebeam$ distributions for the
reconstructed $B$-meson candidates.  Here, $\pvecB$ and $E^*_B$ are the
c.m.~momentum and energy of the $B$ candidate, and \Ebeam is the c.m.~beam
energy.  The simultaneous fit to all three decay modes, assuming
isospin symmetry\footnote{It should be noted that the individual
fit results are in marginal
agreement with the isospin relation.  From Monte Carlo
pseudo-experiments assuming the isospin relation, the authors of the
Belle paper find that the
probability to observe a deviation equal to or larger than the
measurement is 4.9\%.} yields the final Belle result:
\begin{equation}
\BR(\brog) = 1.32\PM 0.34 0.31 \PM 0.10 0.09 \,\,. \nonumber \\
\end{equation}

\begin{figure}[h]
\centering
\includegraphics[width=80mm]{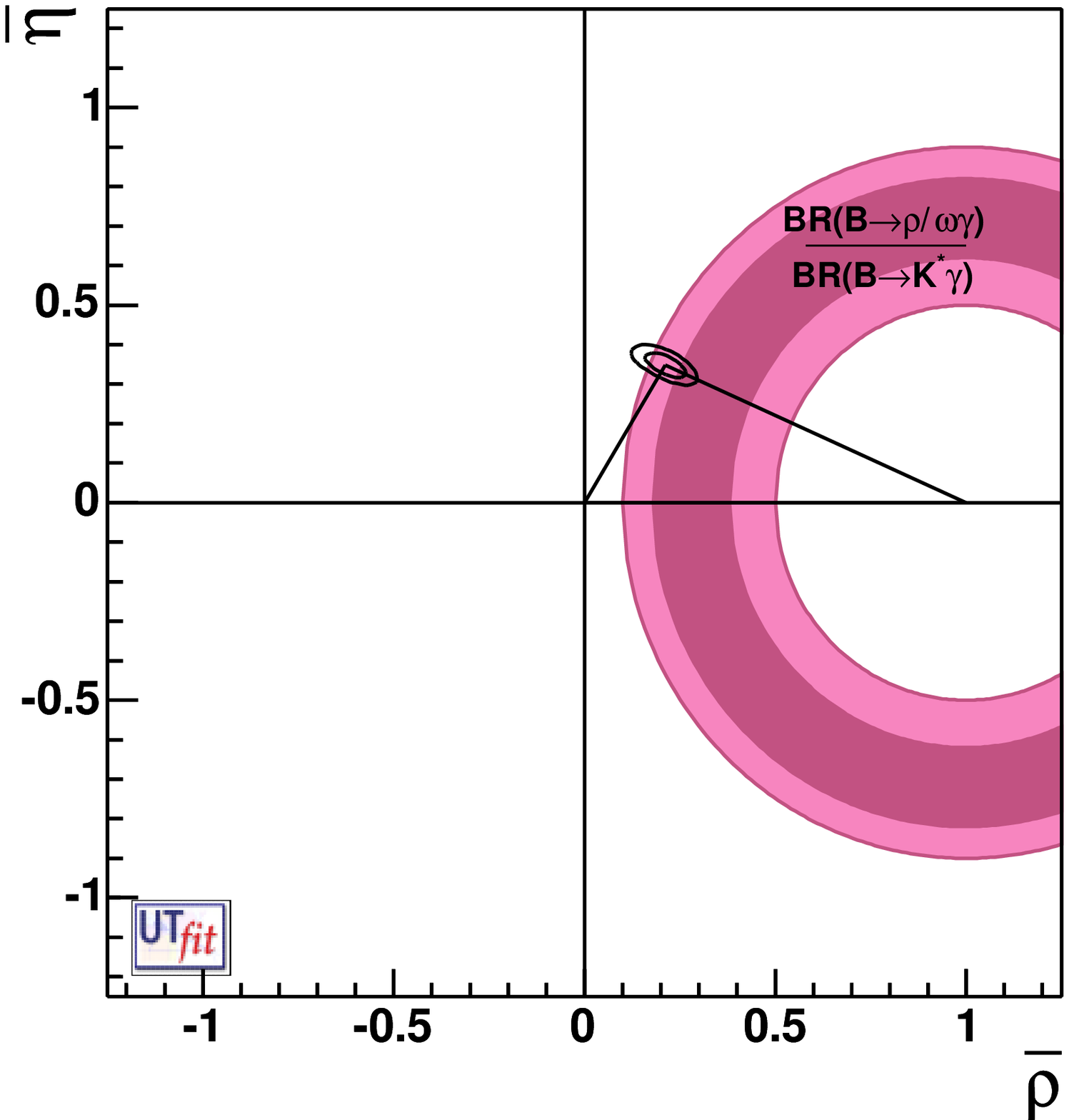}
\caption{The bound on the $(\bar{\rho},\bar{\eta})$ plane from the fit
to the ratio $\BR(\brog)/\BR(\bksg)$~\cite{bib:UTfit}.}
\label{fig:utfit}
\end{figure}

The UTfit Collaboration has performed a fit to the data from the $B$ Factories to
extract $|V_{td}|/|V_{ts}|$ from the ratio $\BR(\brog)/\BR(\bksg)$.  To
represent the theory uncertainties in the ratio of the QCD factorization
expressions for $\brog$ and $\bksg$ decays and in the ratio of the form
factors, the fit assumes flat distributions~\cite{bib:UTfit}.  The result
of the fit is given in Eq.~\ref{eq:vtdvts} and the bound on the
$(\bar{\rho},\bar{\eta})$ plane is shown in Fig.~\ref{fig:utfit}.
\begin{equation}
|V_{td}|/|V_{ts}| = 0.16 \pm 0.02 \,\,. \\
\label{eq:vtdvts}
\end{equation}
The most recent value of this ratio obtained from the $\Delta m_s$
measurement by the CDF Collaboration is about two standard deviations
away from this result~\cite{bib:CDFdelms}.  However, it can be
argued that the non-penguin
contribution to the charged decay mode \brpg\ is significant.  If the
fit is repeated only with the neutral mode, the new result becomes
$|V_{td}|/|V_{ts}|=0.16\pm0.04$.

\section{\boldmath\bsll}

\subsection{Inclusive \boldmath\bsll}
In addition to the photon penguin diagram of \bsg, at the lowest-order
\bsll\ process has contributions from two other diagrams in the SM: a
weak penguin diagram, where the $Z$-boson replaces the photon and a $W$-box
diagram.  Because of these contributions, OPE for \bsll\ has ten 
operators (not counting the operators that are suppressed by increasing
orders of $1/m_W$). The investigation of \bsll\ can identify the sign of the Wilson
coefficient $C_7$ of the photon penguin operator, and can provide the values for the
Wilson coefficients $C_9$ and $C_{10}$ of the vector and axial-vector
electroweak operators.

The inclusive \bsll\ rate has been measured by \babar\ and Belle, using a
sum-of-exlusive-states (semi-inclusive) method.  (So far, unlike \bsg, there
has not been any fully-inclusive measurements.)  The \babar\ measurement, based
on 82\invfb of data, reconstructs 10 hadronic final states, whereas the Belle
measurement utilizes 140\invfb of data and reconstructs 18 hadronic final
states~\cite{bib:exp_xsll}.
The results of these measurements, given as a function of the dilepton mass
squared $q^2$, have been compared with the next-to-next-to-leading-log
predictions for the two choices of the sign of $C_7$.  As listed in
Table~\ref{tab:xsll}, the sign of $C_7$ is quite unlikely to be different
from that in the SM~\cite{bib:theory_xsll}.
\begin{table}[h]
\begin{center}
\caption{\babar\ and Belle measurements for the inclusive branching fraction
(in units of $10^{-6}$) of \bxsll\ in two ranges of the
dilepton invariant mass squared.  The
weighted average is compared to the predictions in the SM and with reversed
sign of $C_7$~\cite{bib:theory_xsll}.}
\begin{tabular}{|l|c|c|} \hline
$q^2$ range  & $1-6\gevns^2$  & $>4m_{\mu}^2$  \\ \hline\hline
\babar\  & $1.8\pm0.9$ & $5.6\pm2.0$ \\
Belle    & $1.5\pm0.6$ & $4.1\pm1.1$ \\ \hline
Average  & $1.6\pm0.5$ & $4.5\pm1.0$ \\ \hline
SM       & $1.6\pm0.2$ & $4.4\pm0.7$ \\
$C_7\to-C_7$ & $3.3\pm0.3$ & $8.8\pm1.0$ \\ \hline
\end{tabular}
\label{tab:xsll}
\end{center}
\end{table}

\subsection{Exclusive \boldmath\bsll}

\begin{figure}[h]
\centering
\includegraphics[width=80mm]{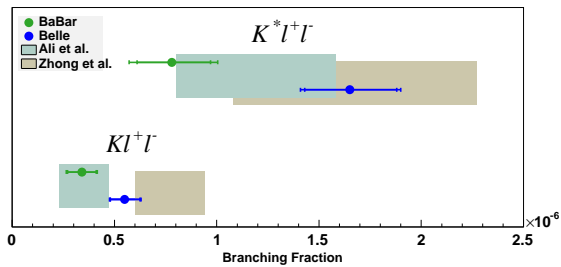}
\caption{Comparison of Belle~\cite{bib:belle_kll} and
\babar\ measurements~\cite{bib:babar_kll} of the
\bkll\ and \bksll\ branching fractions with two predictions
from the SM, using different form factor calculations~\cite{bib:kllbftheory}.
(The plot is courtesy of J.~Berryhill.)}
\label{fig:kllbf}
\end{figure}

Two exclusive \bsll\ decays have been studied extensively at the
$B$ Factories:  \bkll\ and \bksll, where $\ell=e,\mu$.  While their
branching fractions (Fig.~\ref{fig:kllbf}) are comparable to the
recently observed exclusive \bdg\ decays,\footnote{Note that \bkll\ is
currently the lowest-branching-fraction $B$ decay measured at better
than $5\sigma$ significance.} they are experimentally cleaner and
it is possible to measure ratios and asymmetries in the rates, which
can usually be predicted more reliably than the branching fractions.

\begin{figure}[h]
\centering
\includegraphics[width=80mm]{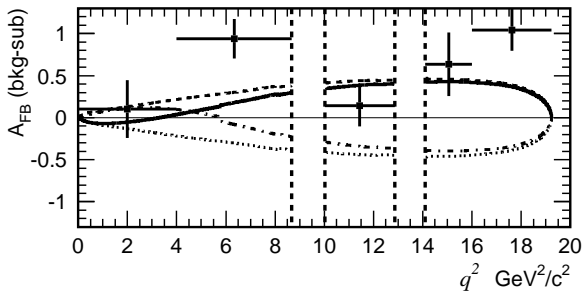}
\caption{Forward-backward asymmetry in \bksll\ decays as measured
by the Belle Collaboration in five bins of $q^2$~\cite{bib:belle_afb}.
The fit result for negative $A_7$ solution (solid) is shown along
with some alternative choices for the signs and values of the
Wilson coefficients; $A_7$ positive case (dashed), $A_{10}$ positive
case (dot-dashed), or both $A_7$ and $A_{10}$ positive case. 
}
\label{fig:belle_afb}
\end{figure}

Of particular interest is the forward-backward asymmetry, \Afb, defined for
the angle of the positively charged lepton with respect to the flight
direction of the \Bz or \Bp meson in the dilepton rest frame.  In \bksll\ decay,
the SM predicts a distinctive pattern for this quantity as a function of $q^2$,
but the presence of new physics can alter its sign and magnitude
dramatically.  A fit to \Afb\ can be used to extract the sign of $C_7$ and
the values of $C_9$ and $C_{10}$.

Figure~\ref{fig:belle_afb} shows the \Afb\ measurement from the Belle
Collaboration, in five bins of $q^2$.  The plot also shows the projection
of the fit that has been performed by fixing the value of $A_7$ to its
SM value of $-0.330$ and choosing $A_9/A_7$ and $A_{10}/A_7$ as fit
parameters, where $A_i$ are the leading terms in $C_i$.  Not shown is the
alternative fit in which $A_7$ is set to its sign-flipped value.  The
results of both fits, as given in Table~\ref{tab:belle_afb}, are consistent
with the SM predictions.  New physics scenarios with positive $A_9A_{10}$
are excluded at $98.2\%$ confidence.
\begin{table}[h]
\begin{center}
\caption{The ratios of the Wilson coefficients obtained from the
fits to the Belle \Afb\ shape in \bksll\ 
decays~\cite{bib:belle_afb}.  Two independent fits
have been performed, with the coefficient $A_7$ fixed to either its
SM value (negative $A_7$) or fixed to its sign-flipped value.  Also
listed are the SM predictions for the measured ratios.}
\begin{tabular}{|l|c|c|c|} \hline
             &  SM     & fitted value     & fitted value \\
             &  value  & with $A_7=-0.33$ & with $A_7=+0.33$  \\ \hline
$A_9/A_7$    & $-12.3$ & $-15.3\PM 3.4 4.8 \pm1.1$ & $-16.3\PM 3.7 5.7 \pm1.4$ \\
$A_{10}/A_7$ & $~12.8$ & $~10.3\PM 5.2 3.5 \pm1.8$ & $~11.1\PM 6.0 3.9 \pm2.4$ \\ \hline
\end{tabular}
\label{tab:belle_afb}
\end{center}
\end{table}

\begin{figure}[h]
\centering
\includegraphics[width=80mm]{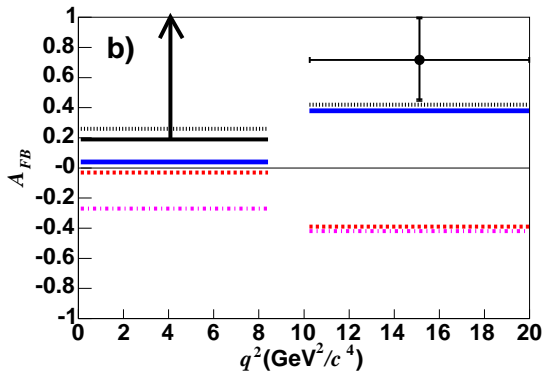}
\caption{Forward-backward asymmetry in \bksll\ decays as measured
by the \babar\ Collaboration in two bins of $q^2$~\cite{bib:babar_kll}.
The arrow in the low-$q^2$ bin indicates the allowed region at 95\% C.L.
The SM prediction (solid) is shown, as are the predictions from
$C_7$ sign-flipped (dotted), $C_9C_{10}$ sign-flipped (dashed) and
both $C_7$ and $C_9C_{10}$ sign-flipped cases.}
\label{fig:babar_afb}
\end{figure}

Figure~\ref{fig:babar_afb} shows the results of the same measurement
from the \babar\ Collaboration.  Similar to the Belle results, the
large asymmetry observed in the high-$q^2$ range disfavors new physics
scenarios in which the sign of the product $C_9C_{10}$ is flipped
from its SM value.  For the low-$q^2$ range, a lower limit on the
asymmetry is obtained: $\Afb>0.19\%$ at 95\% C.L.  It should be noted
that the SM prediction for this range is slightly below this lower
limit~\cite{bib:babar_kll}.

In addition to the \Afb, \babar\ Collaboration has also measured the
fraction of the longitudinal polarization, $F_L$, of \Kstar.  $F_L$ is
sensitive to left-handed currents with complex phases or to right-handed
currents in the photon penguin amplitude.  This measurement has
the potential for determining the sign of $C_7$ independently, however
the current value is in agreement with both signs.  With 1\invab of
data, wrong-sign $C_7$ could be excluded.

For \bkll, the \Afb\ prediction is zero for all $q^2$, both in the SM
and many of its extensions.  Measurements from both experiments are
consistent with this prediction.  In this decay, \babar\ has also
attempted to extract the fraction of (pseudo-)scalar contribution
to the decay amplitude, however with the limited statistics the result
is inconclusive.  With more data, the developed technique will prove
to be useful.

\section{Conclusion}

Radiative penguins have been one of the cornerstones of the search
for new physics in the last decade.  While the current measurements
have not indicated deviations from the Standard Model, the data
collected at the $B$ Factories have significantly improved the
experimental precision, putting more and more stringent constraints
on the physics beyond the SM.  The Wilson coefficients are now experimental
observables.

The developed techniques are highly scalable.  The measurements summarized
in this review utilize only one-fourth to one-half of the data that is
already available.  Both $B$ Factories continue running with increasing
luminosities and the collected data is likely to be doubled before the
end of their runs.

In addition to the direct role they play in the search for new physics,
radiative penguins also provide highly-valueable auxiliary information.
The heavy quark expansion parameters and the shape function derived from
inclusive \bsg\ has already been used to reduce the uncertainties
in the extraction of $|V_{cb}|$ and $|V_{ub}|$.  The measurement of
the \bdg\ penguin can be considered like the $\Delta m_s$ of $B$ Factories.

\bigskip 
\begin{acknowledgments}
I would like to thank the organizers of the conference for an interesting
and enjoyable conference, Asoka~de~Silva for his input on the typesetting of
this document, and many collegues from the \babar\ Collaboration
for very helpful suggestions during the preparation of the talk.
\end{acknowledgments}

\bigskip 


\begin{thebibliography}{99}
\bibitem{bib:kakuno_fpcp}
Time-dependent $CP$ violation measurements are covered in
H.~Kakuno's talk ``New Physics Searches at B Factories'' at this
conference.
\end{thebibliography}
\end{document}